\documentclass[11pt]{article}
\usepackage{epsfig}
\begin{document}
\input amssym.def 
\input amssym
\hfuzz=5.0pt
%
%
%
%
\def\vec#1{\mathchoice{\mbox{\boldmath$\displaystyle\bf#1$}}
{\mbox{\boldmath$\textstyle\bf#1$}}
{\mbox{\boldmath$\scriptstyle\bf#1$}}
{\mbox{\boldmath$\scriptscriptstyle\bf#1$}}}
\def\mbf#1{{\mathchoice {\hbox{$\rm\textstyle #1$}}
{\hbox{$\rm\textstyle #1$}} {\hbox{$\rm\scriptstyle #1$}}
{\hbox{$\rm\scriptscriptstyle #1$}}}}
\def\operatorname#1{{\mathchoice{\rm #1}{\rm #1}{\rm #1}{\rm #1}}}
\chardef\ii="10
\def\widehat{\mathaccent"0362 }
\def\widetilde{\mathaccent"0365 }
\def\vphi{\varphi}
\def\vrho{\varrho}
\def\vtheta{\vartheta}
\def\ih{{\i\over\hbar}}
\def\hi{\frac{\hbar}{\i}}
\def\CD{{\cal D}}
\def\CE{{\cal E}}
\def\CH{{\cal H}}
\def\CL{{\cal L}}
\def\CP{{\cal P}}
\def\CV{{\cal V}}
\def\half{{1\over2}}
\def\bhalf{\hbox{$\half$}}
\def\viert{{1\over4}}
\def\bviert{\hbox{$\viert$}}
\def\dfrac#1#2{\frac{\displaystyle #1}{\displaystyle #2}}
\def\intT{\ih\int_0^\infty\d\,T\,e^{\i ET/\hbar}}
\def\pathint#1{\int\limits_{#1(t')=#1'}^{#1(t'')=#1''}\CD #1(t)}
\def\hbarm{{\dfrac{\hbar^2}{2m}}}
\def\ihbar{\dfrac\i\hbar}
\def\intt{\int_{t'}^{t''}}
\def\bbbr{{\rm I\!R}}                                
\def\bbbz{{\mathchoice {\hbox{$\sf\textstyle Z\kern-0.4em Z$}}
{\hbox{$\sf\textstyle Z\kern-0.4em Z$}}
{\hbox{$\sf\scriptstyle Z\kern-0.3em Z$}}
{\hbox{$\sf\scriptscriptstyle Z\kern-0.2em Z$}}}}    
\def\Ai{\operatorname{Ai}} 
\def\dt{\d t}
\def\ds{\d s}
\def\d{\operatorname{d}}
\def\e{\,\operatorname{e}}
\def\i{\operatorname{i}}
\def\cn{\operatorname{cn}}
\def\dn{\operatorname{dn}}
\def\sn{\operatorname{sn}}
\def\vphi{\varphi}
\def\DI{D_{\,\rm I}}
\def\threedDI{D_{\,3d-\rm I}}
\def\DII{D_{\,\rm II}}
\def\threedDII{D_{\,3d-\rm II}}
\def\DIII{D_{\,\rm III}}
\def\DIV{D_{\,\rm IV}}
\def\operatorname#1{{\mathchoice{\rm #1}{\rm #1}{\rm #1}{\rm #1}}}
\def\bbbone{{\mathchoice {\rm 1\mskip-4mu l} {\rm 1\mskip-4mu l}
{\rm 1\mskip-4.5mu l} {\rm 1\mskip-5mu l}}}
\def\pathint#1{\int\limits_{#1(t')=#1'}^{#1(t'')=#1''}\CD #1(t)}
\def\pathints#1{\int\limits_{#1(0)=#1'}^{#1(s'')=#1''}\CD #1(s)}

\begin{titlepage}
\centerline{\normalsize DESY 05--221 \hfill ISSN 0418 - 9833}
\vskip.5in
\message{TITLE:}
\begin{center}
{\LARGE Path Integral Approach for Spaces \\[3mm]
of Non-constant Curvature in Three Dimensions}
\end{center}
\message{Path Integral Approach for Spaces of Non-constant Curvature
in Three Dimensions}
\vskip.5in
\begin{center}
{\Large Christian Grosche}
\vskip.2in
{\normalsize\em II.\,Institut f\"ur Theoretische Physik}
\vskip.05in
{\normalsize\em Universit\"at Hamburg, Luruper Chaussee 149}
\vskip.05in
{\normalsize\em 22761 Hamburg, Germany}
\end{center}
\normalsize
\vskip.5in
\begin{quote}
Contribution to the ``II. International Workshop on Superintegrable Systems 
in \linebreak 
Classical and Quantum Mechanics'', Dubna, Russia, June 27 - July 1, 2005. 
\end{quote}
\vfill
\begin{center}
{\bf Abstract}
\end{center}
In this contribution I show that it is possible to construct
three-dimensional spaces of non-constant curvature, i.e. three-dimensional
Darboux-spaces. Two-dimensional Darboux spaces have been
introduced by Kalnins et al., with a path integral approach by the present
author. In comparison to two dimensions, in three dimensions it is necessary 
to add a curvature term in the Lagrangian in order that the quantum motion
can be properly defined. Once this is done, it turns out that in the 
two three-dimensional Darboux spaces, which are discussed in this paper,
the quantum motion is similar to the two-dimensional case.
In $\threedDI$ we find seven coordinate systems which separate the 
Schr\"odinger equation. For the second space, $\threedDII$, all coordinate
systems  of flat three-dimensional Euclidean space which separate the
Schr\"odinger equation also separate the Schr\"odinger equation in
$\threedDII$. I solve the path integral on $\threedDI$ in the
$(u,v,w)$-system, and on $\threedDII$ in the
$(u,v,w)$-system and in spherical coordinates.
\end{titlepage}
 
\setcounter{page}{1}
\setcounter{equation}{0}
\section{Introduction}
\message{Introduction}
In this paper the quantum motion on three-dimensional spaces of non-constant
curvature is studied. In \cite{KalninsKMWinter,KalninsKWinter} two-dimensional
spaces of non-constant curvature, called Darboux spaces, were
introduced. Particular emphasis was put on separation of variables
and to find all coordinate systems which separate the Schr\"odinger equation
(respectively the Helmholtz equation) and the path integral. Another important
issue was to find all potentials in these spaces which are
superintegrable. These potentials have the property that there are additional
constants of motion and that the corresponding Schr\"odinger equation separates
in more that one coordinate system. Actually, in two dimensions these systems
have three constants of motion.

In \cite{GROas} the path integral method \cite{FEYb,GRSh,KLEo,SCHUHd} was 
applied to study the free motion on the four Darboux spaces $\DI$--$\DIV$, 
and a study of superintegrable potentials was completed in~\cite{GROPOe}.

In \cite{KalninsKMWinter,KalninsKWinter} the two-dimensional Darboux
were introduced as follows (we also insert for the coordinates $x=u+\i
v,y=u-\i v$, and $(u,v)$ will be called the $(u,v)$-system):
\begin{eqnarray}
\!\!\!\!\!\!\!\!
({\rm I})  \qquad  \d s^2&=& (x+y)\d x\d y
=2u(\d u^2+ \d v^2)
\label{DarbouxI}
\\[2mm]
\!\!\!\!\!\!\!\!
({\rm II}) \qquad \d s^2&=& \bigg(\frac{a}{(x-y)^2}+b\bigg) \d x\d y
=\frac{bu^2-a}{u^2}(\d u^2+\d v^2)
\label{DarbouxII}
\\[2mm]
\!\!\!\!\!\!\!\!
({\rm III})\qquad \d s^2&=& \big(a\,\e^{-(x+y)/2}+b\,\e^{-x-y}\big)
\d x\d y
=\e^{-2u}(b+a\,\e^u)(\d u^2+\d v^2)
\label{DarbouxIII}
\\[2mm]
\!\!\!\!\!\!\!\!
({\rm IV}) \qquad \d s^2&=& -\frac{a\big(\e^{(x-y)/2}+\e^{(y-x)/2}\big)+b}
{\big(\e^{(x-y)/2}-\e^{(y-x)/2}\big)^2}\d x\d y
=\left(\frac{a_+}{\sin^2u}+\frac{a_-}{\cos^2u}\right)(\d u^2+\d v^2)\enspace.
\label{DarbouxIV}
\end{eqnarray}
$a$ and $b$ are additional (real) parameters ($a_\pm=(a\pm 2b)/4$). 
$\DII$ has the property that for $a=0$, $b=1$, we recover the
two-dimensional Euclidean plane, and all four coordinate systems on
the two-dimensional Euclidean plane are also separable  coordinate
systems on $\DII$ for the Schr\"odinger, respectively the Helmholtz equation.
For $b=0$, $a=-1$ the two-dimensional hyperboloid is contained as a second
special case. 

Let us consider three-dimensional generalization of the Darboux spaces
$\DI,\DII$, respectively, with the following line elements 
\begin{eqnarray}
({\rm I})  \qquad  \d s^2&=&2u(\d u^2+\d v^2+\d w^2)\enspace,
\label{DarbouxI-3D}
      \\
({\rm II})  \qquad  \d s^2&=&\frac{bu^2-a}{u^2}(\d u^2+\d v^2+\d w^2)\enspace,
\label{DarbouxII-3D}
\end{eqnarray} 
and $w$ is the new variable. The present cases of
(\ref{DarbouxI-3D},\ref{DarbouxII-3D}), which we call
three-dimensional Darboux space I and II, for short $\threedDI,\threedDII$, 
respectively, are studied in this contribution.
In comparison to their two-dimensional analogies new features appear.
If we consider the Laplace-Beltrami operator,
$\Delta_{LB}=g^{-1/2}\partial_{q^a} g^{ab}g^{1/2}\partial_{q^b}$,
we see that $g^{ab}g^{1/2}\not=\bbbone$, as it is always the case 
in two dimensions
if the metric tensor is proportional to the unit tensor. We obtain 
an additional term $\propto (g^{ab}\Gamma_a+g^{ab}_{,a})\partial_{q^b}$, where
$\Gamma_a=\partial_{q^a}\ln\sqrt{\det(g_{ab})}$. This has the consequence that 
curvature terms $\propto\hbar^2$ appear in the quantum Hamiltonian 
which must be dealt with. We will see that we must add such curvature terms 
in the metrics (\ref{DarbouxI-3D},\ref{DarbouxII-3D}) which cancel 
these $\propto\hbar^2$-terms in the quantization procedure. If this is done,
proper quantum systems can be established.

In the following, we study the cases of (\ref{DarbouxI-3D},\ref{DarbouxII-3D})
within the path integral approach, first for (\ref{DarbouxI-3D}) and second for
(\ref{DarbouxII-3D}). In both cases we find the coordinate systems
which separate the Schr\"odinger equation, respectively the path integral.
We solve the path integral in each space in the $(u,v,w)$-system, and the path
integral corresponding to (\ref{DarbouxII-3D}) also in spherical coordinates.
The extension to this study to find the coordinate systems in the two other
three-dimensional Darboux-spaces and to find the corresponding path integral
solutions will be subject to a future publication.
The last section contains a summary of the achieved results and an outlook.


\setcounter{equation}{0}
\section{The Path Integral Solution \\on the Three-Dimensional Darboux Space
   $\threedDI$}
\message{The Path Integral Solution on the Three-Dimensional Darboux Space
  3D-DI}
We start with the three-dimensional Darboux space $\threedDI$ and
consider the metric:
\begin{equation}
\d s^2=2u(\d u^2+\d v^2+\d w^2)\enspace.
\end{equation} 
The proper definition of the range of the variables $(u,v,w)$ depends
on the proper definition of the space we in fact consider.
As it is known from the two-dimensional case \cite{KalninsKWinter}, 
an embedding in
a three-dimensional Euclidean space yields $a>\half$, whereas an 
embedding in a three-dimensional Minkowskian space yields $a>0$.
We assume in the following that $u>a$, where $a>0$ and that there is no
restriction on the variables $v,w$. They can be cyclic or range within
the entire real line.
According to the general theory we have $g=\det(g_{ab})=(2u)^3$,
therefore $\Gamma_u=3/2u,\Gamma_v=\Gamma_w=0$. 
The Laplace-Beltrami operator has the form
\begin{equation}
\Delta_{LB}=\frac{1}{2u}\bigg(\frac{\partial^2}{\partial u^2}
-\frac{1}{2u}\frac{\partial}{\partial u}
+\frac{\partial^2}{\partial v^2}
+\frac{\partial^2}{\partial w^2}\bigg),\qquad
p_u=\hi\bigg(\frac{\partial}{\partial u}+\frac{3}{4u}\bigg)\enspace,
\end{equation}
and $p_u$ are the corresponding momentum operator for the coordinate $u$.
Of course, $p_v=\hi\partial_v,p_w=\hi\partial_w$. According to our theory we
can calculate the corresponding quantum potential by means of \cite{GRSh}
\begin{equation}
\Delta V=\frac{\hbar^2}{8m}\frac{D-2}{f^4}
\Big[(D-4){f'}^2+2ff''\Big]\enspace,
\end{equation}
provided the metric is proportional to the unit tensor
$(g_{ab})=f^2\bbbone_3$. Indeed $f=\sqrt{2u}$ and $D=3$, which yields
\begin{equation}
\Delta V=-\frac{3\hbar^2}{64mu^3}\enspace.
\label{DeltaV3dDI}
\end{equation}
This gives an effective Lagrangian in the corresponding 
path integral in the product form definition
 \cite{GRSh}
\begin{equation}
\CL_{\rm eff}(u,\dot u,v,\dot v,w,\dot w)
=\frac{m}{2}(2u)(\dot u^2+\dot v^2+\dot w^2)+\frac{3\hbar^2}{64mu^3}
\enspace.
\end{equation}
The quantum potential (\ref{DeltaV3dDI}) has actually the form
of a Schwarzian derivative. Performing a space-time transformation
in the path integral where $u^2=4r$ cancels $\Delta V$,
and produces in turn in the transformed Lagrangian $\CL_E=\CL_{\rm eff}+E$ 
a potential $\propto 2E/\sqrt{r}$ (coupling constant metamorphosis). 
Potentials like this are called ``conditionally solvable'' \cite{GROaf}. 
However, in order that they are in fact conditionally solvable, requires that
an additional potential of the form of $\Delta V$ is 
present. This is not the case here after the transformation into the new
variable $r$ and the corresponding time-transformation; we are left with 
an intractable path integral.

In order to obtain a proper quantum theory on $\threedDI$, we therefore
{\it define} our quantum theory for the free motion on $\threedDI$ as follows:
\begin{eqnarray}
H_{\threedDI}&=&-\hbarm\frac{1}{2u}\bigg(
\frac{\partial^2}{\partial u^2}-\frac{1}{2u}\frac{\partial^2}{\partial u}
+\frac{\partial^2}{\partial v^2}+\frac{\partial^2}{\partial w^2}
\bigg)+\frac{3\hbar^2}{64mu^3}
\nonumber\\   &&
=\frac{1}{2m}\frac{1}{\sqrt{2u}}(p_u^2+p_v^2+p_w^2)\frac{1}{\sqrt{2u}}\enspace.
\end{eqnarray}
This gives in turn a proper definition of the Lagrangian on $\threedDI$
\begin{equation}
\CL^{(\threedDI)}_{\rm eff}(u,\dot u,v,\dot v,w,\dot w)
:=\frac{m}{2}(2u)(\dot u^2+\dot v^2+\dot w^2)\enspace,
\end{equation}
and the path integral has the form
\begin{eqnarray}
&&\!\!\!\!\!\!\!\!\!\!
K^{(\threedDI)}(u'',u',v'',v',w'',w';T)
\nonumber\\   &&\!\!\!\!\!\!\!\!\!\!
:=\lim_{N\to\infty}\bigg(\frac{m}{2\pi\i\epsilon\hbar}\bigg)^N
\prod_{j=1}^{N-1}\int (2u_j)^{3/2}\d u_j\d v_j\d w_j
\exp\Bigg[\frac{\i m}{\hbar}\sum_{j=1}^N
     \widehat{u_j}(\Delta^2u_j+\Delta^2v_j+\Delta^2w_j)\Bigg]
\nonumber\\   &&\!\!\!\!\!\!\!\!\!\!
=\pathint{u}\pathint{v}\pathint{w}(2u)^{3/2}
 \exp\Bigg[\frac{\i m}{\hbar}\int_0^Tu(\dot u^2+\dot v^2+\dot w^2)\dt\Bigg].
\qquad
\label{K3dDI}
\end{eqnarray}
($u_j=u(t_j)$, $\Delta u_j=u_j-u_{j-1}$, $\epsilon=T/N$, 
$\widehat{u_j}=\sqrt{u_{j-1}u_j}$). One may now ask, why we subtract in the
definition of our quantum theory on $\threedDI$ a quantum potential of order
$\hbar^2$, which is usually {\it absolutely necessary to incorporate} 
\cite{GRSh}. It is well-known that other lattice definitions lead to other
quantum potentials which in turn correspond to different ordering
prescriptions in the quantum Hamiltonian \cite{GRSh}. 
This problem can be addresses as follows:
If we perform a time-transformation in the path integral (\ref{K3dDI}), it is 
inevitable that we switch in this procedure to a lattice as given in 
(\ref{K3dDI}) \cite{KLEo}. Had we set up our path integral in a different
lattice as in (\ref{K3dDI}) corresponding to another ordering prescription for
the quantum Hamiltonian $H_{\threedDI}$, say the midpoint prescription and
Weyl-ordering, respectively, we must switch to the lattice in 
(\ref{K3dDI}) in order to perform the time-transformation properly. This
changing of the lattice in turn would produce additional quantum terms 
of order $\hbar^2$ which then would lead back to our definition (\ref{K3dDI}).

From Table \ref{tableE3}\ we can determine the coordinate systems of three
dimensional Euclidean space which separate the Schr\"odinger equation 
for the quantum
motion in $\threedDI$, respectively the path integral (\ref{K3dDI}).
We find the  Cartesian, the three circular systems, the parabolic and the
paraboloidal systems (we take $u=z$ and $(v,w)=(y,z)$), and in addition the 
rotated $(r,q)$-system from \cite{KalninsKWinter} with the additional variable
$w$. This gives seven coordinate systems for $\threedDI$. 
We discuss only the first, the $(u,v,w)$-system:
A rotated system is very similar to the $(u,v,w)$-system;
the three circular systems are contained in the $(v,w)$-coordinates
as subsystems; for the parabolic and the paraboloidal systems which separate
the Schr\"odinger equation, however, we encounter intractable power-terms
similar as  in~\cite{GROas}. 

In the path integral (\ref{K3dDI}) we perform a time
transformation according to $\Delta t_{(j)}=2\widehat{u_j}\Delta
s_{(j)}$, i.e. with time-transformation function $f^2(u)=2u$, and we get: 

\newpage\noindent%
\begin{table}[t!]
\caption{\label{tableE3} Coordinates in Three-Dimensional Euclidean Space}
\hfuzz=15pt
\begin{eqnarray}\begin{array}{l}\vbox{\small\offinterlineskip
\halign{&\vrule#&$\strut\ \hfil\hbox{#}\hfill\ $\cr
\noalign{\hrule}
height2pt&\omit&&\omit&\cr
&Coordinate System &&Coordinates                                  &\cr
height2pt&\omit&&\omit&\cr
\noalign{\hrule}\noalign{\hrule}
height2pt&\omit&&\omit&\cr
&I.~Cartesian &&$x=x'$, $y=y'$,  $z=z'$                           &\cr
height2pt&\omit&&\omit&\cr
\noalign{\hrule}height2pt&\omit&&\omit&\cr
&II.~Circular Polar
              &&$x=\vrho\cos\vphi$, $y=\vrho\sin\vphi$, $z=z'$            &\cr
height2pt&\omit&&\omit&\cr
\noalign{\hrule}height2pt&\omit&&\omit&\cr
&III.~Circular Elliptic
              &&$x=d\cosh\mu\cos\nu$, $y=d\sinh\mu\sin\nu$,$z=z'$ &\cr
height2pt&\omit&&\omit&\cr
\noalign{\hrule}height2pt&\omit&&\omit&\cr
&IV.~Circular Parabolic
              &&$x=\half(\eta^2-\xi^2)$, $y=\xi\eta$, $z=z'$      &\cr
height2pt&\omit&&\omit&\cr
\noalign{\hrule}height2pt&\omit&&\omit&\cr
&V.~Sphero-Conical
              &&$x=r\sn(\alpha,k)\dn(\beta,k')$, 
                $y=r\cn(\alpha,k)\cn(\beta,k')$&\cr 
&             &&$z=r\dn(\alpha,k)\sn(\beta,k')$&\cr
height2pt&\omit&&\omit&\cr
\noalign{\hrule}height2pt&\omit&&\omit&\cr
&VI.~Spherical&&$x=r\sin\vtheta\cos\vphi$, 
                $y=r\sin\vtheta\sin\vphi$, 
                $z=r\cos\vtheta$               &\cr
height2pt&\omit&&\omit&\cr
\noalign{\hrule}height2pt&\omit&&\omit&\cr
&VII.~Parabolic&&$x=\xi\eta\cos\vphi$, 
                 $y=\xi\eta\sin\vphi$, 
                 $z=\half(\eta^2-\xi^2)$        &\cr
height2pt&\omit&&\omit&\cr
\noalign{\hrule}height2pt&\omit&&\omit&\cr
&VIII.~Prolate Spheroidal
              &&$x=d\sinh\mu\sin\nu\cos\vphi$, 
                $y=d\sinh\mu\sin\nu\sin\vphi$  &\cr
&             &&$z=d\cosh\mu\cos\nu$           &\cr
height2pt&\omit&&\omit&\cr
\noalign{\hrule}height2pt&\omit&&\omit&\cr
&IX.~Oblate Spheroidal
              &&$x=d\cosh\mu\sin\nu\sin\vphi$, 
                $y=d\cosh\mu\sin\nu\sin\vphi$  &\cr
&             &&$z=d\sinh\mu\cos\nu$           &\cr
height2pt&\omit&&\omit&\cr
\noalign{\hrule}height2pt&\omit&&\omit&\cr
&X.~Ellipsoidal&&$x=k^2\sqrt{a^2-c^2}\sn\alpha\sn\beta\sn\gamma$
                                                           &\cr
&   &&$y=-(k^2/k')\sqrt{a^2-c^2}\cn\alpha\cn\beta\cn\gamma$&\cr
&   &&$z=(\i/k')\sqrt{a^2-c^2}\dn\alpha\dn\beta\dn\gamma$  &\cr
height2pt&\omit&&\omit&\cr
\noalign{\hrule}height2pt&\omit&&\omit&\cr
&XI.~Paraboloidal&&$x=2d\cosh\alpha\cos\beta\sinh\gamma$, 
      $y=2d\sinh\alpha\sin\beta\cosh\gamma$                &\cr
&   &&$z=d(\cosh^2\alpha+\cos^2\beta-\cosh^2\gamma)$       &\cr
height2pt&\omit&&\omit&\cr
\noalign{\hrule}}}\end{array}\nonumber\end{eqnarray}
\vspace{-1cm}
\end{table}
\hfuzz=5pt
 
\begin{eqnarray}
&&\!\!\!\!\!\!\!\!
K^{(\threedDI)}(u'',u',v'',v',w'',w';T)
\nonumber\\ &&\!\!\!\!\!\!\!\!
=(4u'u'')^{-1/2}
\int_{-\infty}^\infty\frac{\d E}{2\pi\hbar}\,\e^{-\i ET/\hbar}
   \int_0^\infty \d s'' K^{(\threedDI)}(u'',u',v'',v',w'',w';s'')
\\
&&\!\!\!\!\!\!\!\!
\hbox{with $K(s'')$ given by:}\nonumber\\
&&\!\!\!\!\!\!\!\!
K^{(\threedDI)}(u'',u',v'',v',w'',w';s'')
\nonumber\\ &&\!\!\!\!\!\!\!\!
=\pathints{u}\pathints{v}\pathints{w}\exp\left\{\ihbar\int_0^{s''}\bigg[
\frac{m}{2}(\dot u^2+\dot v^2+\dot w^2)+2uE\bigg]\d s\right\}
\nonumber\\ &&\!\!\!\!\!\!\!\!
=\sum_{l_v,l_w=-\infty}^\infty
 \frac{\e^{\i l_v(v''-v')+\i l_w(w''-w')}}{(2\pi)^2}
\exp\bigg(-\ihbar\frac{\hbar^2}{2m}(l_v^2+l_w^2)s''\bigg)
\nonumber\\ &&\!\!\!\!\!\!\!\!\qquad\qquad\times
\pathints{u}\exp\left[\ihbar\int_0^{s''}\bigg(
\frac{m}{2}\dot u^2+2uE\bigg)\d s\right]\enspace.
\label{Kuvs}
\end{eqnarray}
I have separated the $(v,w)$-dependent parts of the path integral in
circular waves. Depending on the boundary conditions, also plane
waves can be possible \cite{KalninsKWinter}.
The remaining path integral in the variable $u$ 
is a path integral for the linear potential. 
In a similar way as in \cite{GROas} we obtain for the 
kernel $K(T)$:
\begin{eqnarray}
&&\!\!\!\!\!\!\!\!\!\!
K^{(\threedDI)}(u'',u',v'',v',w'',w';T)
=\int_{-\infty}^\infty\frac{\d E}{2\pi\hbar}\,\e^{-\i ET/\hbar}
\nonumber\\ &&\!\!\!\!\!\!\!\!\!\!\qquad\qquad\qquad\times
\sum_{l_v,l_w=-\infty}^\infty
 \frac{\e^{\i l_v(v''-v')+\i l_w(w''-w')}}{(2\pi)^2}\,
G_u^{(\threedDI)}\bigg(E;u'',u';-\frac{\hbar^2}{2m}\vec L^2\bigg)\enspace,
\end{eqnarray}
and we have abbreviated $\vec L^2=l_v^2+l_w^2$.
For the complete solution we must know the kernel $G_u(u'',u';\CE)$,
which is obtained from the Green  function for the
linear potential $V(x)=kx$, and is given by \cite{GRSh} 
\begin{eqnarray}
G^{(k)}(x'',x';\CE)&=&\frac{4m}{3\hbar}
\bigg[\bigg(x'-\frac{\CE}{k}\bigg)\bigg(x''-\frac{\CE}{k}\bigg)\bigg]^{1/2}
\nonumber\\ &&\qquad\times
I_{1/3}\left[\frac{\sqrt{8mk}}{3\hbar}
\bigg(x_<-\frac{\CE}{k}\bigg)^{3/2}\right]
K_{1/3}\left[\frac{\sqrt{8mk}}{3\hbar}
\bigg(x_>-\frac{\CE}{k}\bigg)^{3/2}\right]\enspace.\quad
\end{eqnarray}
$I_\nu$ and $K_\nu$ are modified Bessel-functions \cite{GRA}, and $x_<$ and 
$x_>$ denote the smaller and larger of $x'$ and $x''$, respectively.
We have to identify $\CE=-\vec L^2\hbar^2/2m$, $k=-2E$, and $x=u$.
In addition, we have to recall that the motion in $u$ takes place only
in the half-space $u>a$. In order to construct
the Green  function in the half-space $x>a$ we have to put Dirichlet
boundary-conditions at $x=a$ \cite{GROr}. Therefore we obtain finally:
\begin{eqnarray}
&&G^{(\threedDI)}(u'',u',v'',v',w'',w';E)
\nonumber\\  &&\qquad
=\sum_{l_v,l_w=-\infty}^\infty
 \frac{\e^{\i l_v(v''-v')+\i l_w(w''-w')}}{(2\pi)^2}
\frac{4m}{3\hbar}(4u'u'')^{-1/2}
\bigg[\bigg(u'-\frac{\vec L^2\hbar^2}{4mE}\bigg)
\bigg(u''-\frac{\vec L^2\hbar^2}{4mE}\bigg)\bigg]^{1/2}
\nonumber\\  &&\qquad\qquad\times
\left[\tilde I_{1/3}\bigg(u_<-\frac{\vec L^2\hbar^2}{4mE}\bigg)
      \tilde K_{1/3}\bigg(u_>-\frac{\vec L^2\hbar^2}{4mE}\bigg)
\vphantom{\dfrac{\tilde I_{1/3}\bigg(a-\frac{\vec L^2\hbar^2}{4mE}\bigg)}
      {\tilde K_{1/3}\bigg(a-\frac{\vec L^2\hbar^2}{4mE}\bigg) }}\right. 
\nonumber\\  &&\qquad\qquad\qquad\qquad\left.
-\dfrac{\tilde I_{1/3}\bigg(a-\frac{\vec L^2\hbar^2}{4mE}\bigg)}
      {\tilde K_{1/3}\bigg(a-\frac{\vec L^2\hbar^2}{4mE}\bigg) }
\tilde K_{1/3}\bigg(u'-\frac{\vec L^2\hbar^2}{4mE}\bigg)
\tilde K_{1/3}\bigg(u''-\frac{\vec L^2\hbar^2}{4mE}\bigg)\right]
\,.\qquad\qquad
\label{G-DarbouxI}
\end{eqnarray}
$\tilde I_\nu(z)$ denotes $\tilde I_\nu(z)=I_\nu\bigg(
\frac{4\sqrt{-mE}}{3\hbar}z^{3/2}\bigg)$, with $\tilde K_\nu(z)$ similarly. 
Due to the relation to the Airy-function \cite{ABS}
$K_{\pm 1/3}(\zeta)=\pi\sqrt{3/z}\, \Ai(z)$, $z=(3\zeta/2)^{2/3}$, and 
the observation that for $E<0$ the argument of $\Ai(z)$ is always greater
than zero, there are no bound states. Let us note that we can replace in 
(\ref{G-DarbouxI}) the expansion of the circular- (respectively
plane-) waves of the $(v,w)$-subsystem by the appropriate expansion
of the remaining three-circular systems of Table \ref{tableE3}, i.e. circular
polar coordinates with a Bessel-function times circular waves, circular
parabolic coordinates with a product of two parabolic cylinder functions and
circular elliptic coordinates with Mathieu-functions \cite{GROad}. 
This concludes the discussion of $\threedDI$.


\setcounter{equation}{0}
\section{The Path Integral Solution\\ on the Three-Dimensional Darboux Space
   $\threedDII$}
\message{The Path Integral Solution on the Three-Dimensional Darboux Space
  3D-DII}
\subsection{The $(u,v,w)$- and the Cylindrical Systems}
For the second three-dimensional Darboux space, we consider the metric
\begin{equation}
\d s^2=\frac{bu^2-a}{u^2}(\d u^2+\d v^2+\d w^2),\qquad
p_u=\hi\bigg(\frac{\partial}{\partial u}+\frac{\Gamma_u}{2}\bigg)
\enspace,
\end{equation} 
and $w$ is the new variable. We can write the metric tensor according to
$(g_{ab})=f^2\bbbone_3$ with $f=h/u$, and $h=\sqrt{bu^2-a}$.
The general theory yields $g=(h/u)^6$, $\Gamma_u=3h'/h-3/u$, and
\begin{eqnarray}
\Delta V&=&\Delta V_1+\Delta V_2
\nonumber\\  
\Delta V_1&=&\frac{\hbar^2}{8mh^6}\Big(2ab(u^2-1)-3b^2u^4\Big),\qquad
\Delta V_2=\frac{3\hbar^2}{8mf^2u^2}\enspace.
\end{eqnarray}
The quantum potential $\Delta V_2$ is necessary in order to obtain the
correct energy spectrum, the quantum potential $\Delta V_1$ is interpreted as
a curvature term which we add in the metric for our proper quantum theory on
$\threedDII$. Therefore similar as for $\threedDI$:
\begin{equation}
\CL^{(\threedDII)}_{\rm eff}(u,\dot u,v,\dot v,w,\dot w)
:=\frac{m}{2}\frac{bu^2-a}{u^2}(\dot u^2+\dot v^2+\dot w^2)
+\Delta V_1\enspace,
\end{equation}
and the quantum Hamiltonian has the form
\begin{eqnarray}
H^{(\threedDII)}&:=&-\frac{\hbar^2}{2m}\frac{u^2}{bu^2-a}
\Bigg[\frac{\partial^2}{\partial u^2}+
\bigg(\frac{b}{f}-\frac{1}{u}\bigg)\frac{\partial}{\partial u}
+\frac{\partial^2}{\partial v^2}+\frac{\partial^2}{\partial w^2}\Bigg]
-\Delta V_1
\\
&=&\frac{1}{2m}\frac{u}{\sqrt{bu^2-a}}(p_u^2+p_v^2+p_w^2)
\frac{u}{\sqrt{bu^2-a}}+\frac{3\hbar^2}{8mf^2u^2}\enspace.
\end{eqnarray}
($p_v,p_w$ as in $\threedDI$).
The special form of the incorporation of a curvature term in $H$ can be 
justified in the same way as in the case for $\threedDI$.
We consider the path integral on $\threedDII$ and obtain by performing a
time-transformation in the usual way:
%
\begin{eqnarray}
&&K^{(\threedDII)}(u'',u',v'',v',w'',w';T)
\nonumber\\  &&
=\pathint{u}\pathint{v}\pathints{w}
\nonumber\\  &&\qquad\times
  \bigg(\frac{bu^2-a}{u^2}\bigg)^{3/2}\exp\Bigg\{\ih\int_0^T\Bigg[
  \frac{m}{2}\frac{bu^2-a}{u^2}(\dot u^2+\dot v^2+\dot w^2)
  -\frac{3\hbar^2}{8mf^2u^2}\Bigg]\dt\Bigg\}
\label{K3dDIIb}\\  &&
=\int_{-\infty}^\infty\frac{\d E}{2\pi\hbar}\,\e^{-\i ET/\hbar}
   [f(u')f(u'')]^{-1/4}\int_0^\infty \d s''
\nonumber\\  &&\qquad\times
   \pathints{u}\pathints{v}\pathints{w}
\nonumber\\  &&\qquad\times
   \exp\left\{\ih\int_0^{s''}\bigg[
   \frac{m}{2}(\dot u^2+\dot v^2+\dot w^2)-\frac{\hbar^2}{2m}
   \frac{2maE/\hbar^2+3/4}{u^2}\bigg]\ds +\ih bEs''\right\}.\qquad\qquad
\label{K3dDII}
\end{eqnarray}
A look on Table \ref{tableE3}\ shows that all eleven coordinate systems
can be used to separate variables in the path integral (\ref{K3dDIIb}).
The path integral in (\ref{K3dDII}) in the variable $u$
can be seen as the special case of $V_1$ (singular oscillator)
from \cite{GROPOa} where $\omega=0,k_{2,3}=\pm\half$. An explicit evaluation
is possible in the $(u,v,w)$-system, in spherical, in circular-polar,
circular-parabolic and in parabolic coordinates. Let us first consider the
$(u,v,w)$-system. We continue in (\ref{K3dDII}) in the same way as in
\cite{GROas}, we set $\lambda^2=1-2m|a|E$, where we assume that
$a<0$, and we get due to the fact that the path integral (\ref{K3dDII})
is of the radial $1/u^2$-type \cite{GRSh}:
\begin{eqnarray}
&&K^{(\threedDII)}(u'',u',v'',v',w'',w';T)
\nonumber\\ &&
=\frac{1}{[f(u')f(u')]^{1/4}}\frac{m\sqrt{u'u''}}{\i\hbar }
\int_{-\infty}^\infty\frac{\d E}{2\pi\hbar}\,\e^{-\i ET/\hbar}
\int_{\bbbr^2}\d \vec k
\frac{\e^{\i k_v(v''-v')+\i k_w(w''-w')}}{(2\pi)^2}
\nonumber\\ &&\qquad\times
\int_0^\infty \frac{\ds''}{s''}
\exp\left[\ih\bigg(bE-\frac{\hbar^2\vec K^2}{2m}\bigg)s''
  +\ih\frac{m}{2s''}({u'}^2+{u''}^2)\right]
  I_\lambda\bigg(\frac{mu'u''}{\i\hbar s''}\bigg)\enspace,
\label{Klambda3dDII}
\end{eqnarray}
where we have set $(\vec K=(k_v,k_w))$ and $\vec K^2=k_v^2+k_w^2$.
The evaluation of the $\ds''$-integration integral in (\ref{Klambda3dDII}) 
yields for the Green function 
\begin{eqnarray}
&&G^{(\threedDII)}(u'',u',v'',v',w'',w';E)
\nonumber\\  &&
=\frac{1}{[f(u')f(u')]^{1/4}}\frac{\hbar}{\pi^2}\int_{\bbbr^2}\d \vec k\,
\frac{\e^{\i k_v(v''-v')+\i k_w(w''-w')}}{(2\pi)^2}
\nonumber\\  &&\qquad\times
\int_0^\infty\frac{2p\sinh\pi p\d p}{\frac{\hbar^2}{2m|a|}(p^2+1)-E}
K_{\i p}\left(\sqrt{\vec K^2-\frac{2mbE}{\hbar^2}}\,u'\right)
K_{\i p}\left(\sqrt{\vec K^2-\frac{2mbE}{\hbar^2}}\,u''\right)\enspace,\qquad
\label{G3dDarbouxIIuv}
\end{eqnarray}
with $\lambda=\sqrt{1-2m|a|E/\hbar^2}\equiv \i p$.
The wave functions and the energy spectrum are read off:
\begin{eqnarray}
\Psi(u,v)&=&\frac{\e^{\i k_v v+\i k_ww}}{2\pi f^{1/4}(u)}
\cdot\frac{\sqrt{2p\sinh\pi p}}{\pi}
K_{\i p}\left(\sqrt{\vec K^2-\frac{2mbE}{\hbar^2}}\,u\right)\enspace,
\\
E&=&\frac{\hbar^2}{2m|a|}(p^2+1)\enspace.
\label{EdreiDII}
\end{eqnarray}
The case $b=0$ gives the case of the quantum motion on the three-dimensional
hyperboloid \cite{GROad}, as it should be. 
Let us note that we can replace in (\ref{G3dDarbouxIIuv}) 
the expansion of the circular- (respectively
plane-) waves of the $(v,w)$-subsystem by the appropriate expansion
of the remaining three circular systems of Table \ref{tableE3}, i.e. 
circular-polar coordinates with a Bessel-function times circular waves, 
circular-parabolic coordinates with a product of two parabolic cylinder 
functions and circular-elliptic coordinates with Mathieu-functions 
\cite{GROad}. 

\subsection{The Spherical System}
We consider the metric on $\threedDII$ in spherical coordinates
\begin{equation}
\d s^2=\bigg(b-\frac{a}{r^2\cos^2\vtheta}\bigg)
(\d r^2+r^2\d\vtheta^2+r^2\sin^2\vtheta\d\vphi^2)
\end{equation}
from which the classical Lagrangian follows
\begin{equation}
\CL(r,\dot r,\vtheta,\dot\vtheta,\vphi,\dot\vphi)=\frac{m}{2}
  \bigg(b-\frac{a}{r^2\cos^2\vtheta}\bigg)
(\dot r^2+r^2\dot\vtheta^2+r^2\sin^2\vtheta^2\dot\vphi^2)\enspace.
\end{equation}
However, as has been pointed out in \cite{GROas} 
this coordinate representation is not very well suited for
our purposes, except that we recover for $a=0$ polar coordinate in $\bbbr^2$. 
We introduce $r=\e^{\tau_2}$ and $\cos\vtheta=1/\cosh\tau_1$.
We also have to take into account the curvature terms in a similar way
as in the previous subsection which means that we obtain in the quantization
procedure one term we subtract and one term we keep. This leads us to the
following definition of the Lagrangian on $\threedDII$ for spherical
coordinates in the transformed ($\tau_1,\tau_2,\vphi)$-system
\begin{eqnarray}
&&\CL^{(\threedDII)}_{\rm eff}
(\tau_1,\dot\tau_1,\tau_2,\dot\tau_2,\vphi,\dot\vphi)
=\frac{m}{2}
\bigg(\frac{b\,\e^{2\tau_2}}{\cosh^2\tau_1}-a\bigg)
(\dot\tau_1^2+\cosh^2\tau_1\dot\tau_2^2+\sinh^2\tau_1\dot\vphi^2)
\nonumber\\  &&\qquad\qquad
-\bigg(\frac{b\,\e^{2\tau_2}}{\cosh^2\tau_1}-a\bigg)^{-1}
\frac{\hbar^2}{2m}\bigg(4+\frac{1}{\cosh^2\tau_1}
-\frac{1}{\sinh^2\tau_1}\bigg).
 \end{eqnarray}
We obtain the following path integral representation
($f(\tau_1,\tau_2)=b\,\e^{2\tau_2}/\cosh^2\tau_1-a$)
\begin{eqnarray}
&&\!\!\!\!\!\!\!\!\!\!\!\!
K^{(\threedDII)}(\tau_1'',\tau_1',\tau_2'',\tau_2',\vphi'',\vphi';T)
\nonumber\\  &&\!\!\!\!\!\!\!\!\!\!\!\!
=\pathint{\tau_1}\pathint{\tau_2}\pathint{\vphi}\sinh\tau_1\cosh\tau_1
\bigg(\frac{b\,\e^{2\tau_2}}{\cosh^2\tau_1}-a\bigg)^{3/2}\qquad\qquad\qquad
\nonumber\\  &&\!\!\!\!\!\!\!\!\!\!\!\!\quad \times
\exp\Bigg\{\ih\int_0^T\Bigg[\frac{m}{2}f(\tau_1,\tau_2)
(\dot\tau_1^2+\cosh^2\tau_1\dot\tau_2^2+\sinh^2\tau_1\dot\vphi^2)
\nonumber\\  &&\!\!\!\!\!\!\!\!\!\!\!\!\qquad\qquad\qquad\qquad
-\frac{1}{f(\tau_1,\tau_2)}\frac{\hbar^2}{2m}\bigg(4+\frac{1}{\cosh^2\tau_1}
-\frac{1}{\sinh^2\tau_1}\bigg)\Bigg]\dt\Bigg\}
\nonumber\\  &&\!\!\!\!\!\!\!\!\!\!\!\!=
[f(\tau_1',\tau_2')f(\tau_1'',\tau_2'')]^{-1/4}
\int_{-\infty}^\infty\frac{\d E}{2\pi\hbar}\,\e^{-\i ET/\hbar}
\int_0^\infty \d s'' 
K^{(\threedDII)}(\tau_1'',\tau_1',\tau_2'',\tau_2',\vphi'',\vphi';s'')
\\  &&\!\!\!\!\!\!\!\!\!\!\!\!\hbox{with the time-transformed path integral 
$K(s'')$ given by  $(a<0)$:}
\nonumber\\
&&\!\!\!\!\!\!\!\! \!\!\!\!
K^{(\threedDII)}(\tau_1'',\tau_1',\tau_2'',\tau_2',\vphi'',\vphi';s'')
\nonumber\\  &&\!\!\!\!\!\!\!\!\!\!\!\!
=\pathints{\tau_1}\pathints{\tau_2}\pathints{\vphi}\sinh\tau_1\cosh\tau_1
\nonumber\\  &&\!\!\!\!\!\!\!\!\!\!\!\!\qquad\times
\exp\Bigg\{\ih\int_0^{s''}\Bigg[
\frac{m}{2}(\dot\tau_1^2+\cosh^2\tau_1\dot\tau_2^2+\sinh^2\tau_1\dot\vphi^2)
\nonumber\\  &&\!\!\!\!\!\!\!\!\!\!\!\!\qquad\qquad\qquad\qquad
+Eb\frac{\e^{2\tau_2}}{\cosh^2\tau_1}-aE
-\frac{\hbar^2}{2m}\bigg(4+\frac{1}{\cosh^2\tau_1}
-\frac{1}{\sinh^2\tau_1}\bigg)\Bigg]\d s\Bigg\}\enspace.
\end{eqnarray}
This path integral in the variable $\vphi$ can be easily evaluated, and in the
variable $\tau_2$ we have a path integral for Liouville quantum mechanics
\cite{GRSh}. This yields
\begin{eqnarray}
&&\!\!\!\!\!\!\!\!\!\!\!\!
K^{(\threedDII)}(\tau_1'',\tau_1',\tau_2'',\tau_2',\vphi'',\vphi';s'')
\nonumber\\  &&\!\!\!\!\!\!\!\!\!\!\!\!
=\sqrt{\cosh\tau_1'\cosh\tau_1'}
\exp\left[\ih \bigg(|a|E-\frac{\hbar^2}{2m}\bigg)s''\right]
\sum_{k_\vphi\in\bbbz}\dfrac{\e^{\i k_\vphi(\vphi''-\vphi')}}{2\pi}
\nonumber\\  &&\!\!\!\!\!\!\!\!\!\!\!\!\qquad\times
\frac{2}{\pi^2}\int_0^\infty\d k_{\tau_2}\,k_{\tau_2}\sinh\pi k_{\tau_2} 
K_{\i k_{\tau_2}}\Bigg(\frac{\sqrt{-2mbE}}{\hbar}\,\e^{\tau_2'}\Bigg)
K_{\i k_{\tau_2}}\Bigg(\frac{\sqrt{-2mbE}}{\hbar}\,\e^{\tau_2''}\Bigg)
\nonumber\\  &&\!\!\!\!\!\!\!\!\!\!\!\!\qquad\times
\pathints{\tau_1}\exp\left\{\ih\int_0^{s''}\Bigg[
\frac{m}{2}\dot\tau_1^2-\frac{\hbar^2}{2m}\bigg(
\frac{k_\vphi^2-\viert}{\sinh^2\tau_1}
-\frac{-k_{\tau_2}^2-\viert}{\cosh^2\tau_1}\Bigg]\d s\right\}
\nonumber\\  &&\!\!\!\!\!\!\!\!\!\!\!\!
=\sqrt{\cosh\tau_1'\cosh\tau_1'}
\exp\left[\ih \bigg(|a|E-\frac{\hbar^2}{2m}\bigg)s''\right]
\sum_{k_\vphi\in\bbbz}\dfrac{\e^{\i k_\vphi(\vphi''-\vphi')}}{2\pi}
\nonumber\\  &&\!\!\!\!\!\!\!\!\!\!\!\!\qquad\times
\frac{2}{\pi^2}\int_0^\infty\d k_{\tau_2}\,k_{\tau_2}\sinh\pi k_{\tau_2} 
K_{\i k_{\tau_2}}\Bigg(\frac{\sqrt{-2mbE}}{\hbar}\,\e^{\tau_2'}\Bigg)
K_{\i k_{\tau_2}}\Bigg(\frac{\sqrt{-2mbE}}{\hbar}\,\e^{\tau_2''}\Bigg)
\nonumber\\  &&\!\!\!\!\!\!\!\!\!\!\!\!\qquad\times
\int_0^\infty\d s''\e^{-\i p^2\hbar s''/2m}
\Psi^{(k_\vphi,\i k_{\tau_2})}(\tau_3'')
\Psi^{(k_\vphi,\i k_{\tau_2})\,*}(\tau_3')
\enspace,\qquad\qquad
\end{eqnarray}
and we have inserted in the last step the path integral solution for the
modified P\"oschl--Teller potential $V^{(mPT)}(r)$. 
The modified  P\"oschl--Teller functions
$\Psi_p^{(\mu,\nu)}(\omega)$  for the continuous spectrum are given by
\cite{BJb,KLEMUS} 
 \begin{eqnarray}
  V^{(mPT)}(r)&=&\hbarm \bigg({\eta^2-{1\over4}\over\sinh^2r}
   -{\nu^2-{1\over4}\over\cosh^2r}\bigg)
  \nonumber\\   
  \Psi_p^{(\eta,\nu)}(r)
  &=&N_p^{(\eta,\nu)}(\cosh r)^{2k_1-\half}(\sinh r)^{2k_2-\half}
  \nonumber\\   &&\qquad\qquad\times
  {_2}F_1(k_1+k_2-\kappa,k_1+k_2+\kappa-1;2k_2;-\sinh^2r)
  \\
  N_p^{(\eta,\nu)}
  &=&{1\over\Gamma(2k_2)}\sqrt{p\sinh\pi p\over2\pi^2}
  \Big[\Gamma(k_1+k_2-\kappa)\Gamma(-k_1+k_2+\kappa)
  \nonumber\\   &&\qquad\qquad\times
  \Gamma(k_1+k_2+\kappa-1)\Gamma(-k_1+k_2-\kappa+1)\Big]^{1/2}\enspace,
\end{eqnarray}
$k_1,k_2$ defined by:
$k_1=\half(1\pm\nu)$, $k_2=\half(1\pm\eta)$, where the correct sign
depends on the boundary-conditions for $r\to0$ and $r\to\infty$,
respectively. The number $N_M$ denotes the maximal number of
states with $0,1,\dots,N_M<k_1-k_2-\half$. $\kappa=k_1-k_2-n$ for the
bound states and $\kappa=\half(1+\i p)$ for the scattering states.
${_2}F_1(a,b;c;z)$ is the hypergeometric function \cite[p.1057]{GRA}.
We omit the bound states, because they do not exist here.

Performing the $s''$-integration gives the energy-spectrum (\ref{EdreiDII})
with the Green function
\begin{eqnarray}
&&G^{(\threedDII)}(\tau_1'',\tau_1',\tau_2'',\tau_2',\vphi'',\vphi';E)
\nonumber\\  &&
=\int_0^\infty \d p\int_0^\infty \d k_{\tau_2}\sum_{k_\vphi\in\bbbz}
\frac{\Psi_{p,k_{\tau_2},k_{\vphi}}(\tau_1'',\tau_2'',\vphi'')
\Psi_{p,k_{\tau_2},k_{\vphi}}^*(\tau_1'',\tau_2'',\vphi'')}
    {\frac{\hbar^2}{2m|a|}(p^2+1)-E}\enspace,
\end{eqnarray}
and the wave-functions are given by
\begin{eqnarray}
&&\!\!\!\!\!\!\!\!\!\!\!\!\!\!\!\!\!\!\!\!\!\!\!\!\!\!
\Psi_{p,k_{\tau_2},k_{\vphi}}(\tau_1,\tau_2,\vphi)
\nonumber\\  &&\!\!\!\!\!\!\!\!\!\!\!\!\!\!\!\!\!\!\!\!\!\!\!\!\!\!=
\frac{\sqrt{2\sinh\tau_1\cosh\tau_1}}{f(\tau_1,\tau_2)^{1/4}}
\dfrac{\e^{\i k_\vphi(\vphi''-\vphi')}}{\sqrt{2\pi}}
\frac{\sqrt{k_{\tau_2}\sinh\pi k_{\tau_2}}}{\pi} \,
K_{\i k_{\tau_2}}\Bigg(\i\sqrt{\frac{b}{|a|}(p^2+1)}\,\e^{\tau_2}\Bigg)
\Psi^{(k_\vphi,\i k_{\tau_2})}(\tau_3),\,
\end{eqnarray}
where the replacement $r=\e^{\tau_2}$ and $\cos\vtheta=1/\cosh\tau_1$
gives the wave-functions in the original spherical system.

Due to the $1/u^2$-term in the metric, it is possible to separate
the path integral in conical coordinates, however, it cannot be evaluated.


\setcounter{equation}{0}
\section{Discussion and Summary}
\message{Discussion and Summary}
Our results are very satisfactory. It was possible to define a quantum theory
on three-dimensional spaces of non-constant curvature and evaluate the
path integral in the $(u,v,w)$-coordinate system in both spaces. In
$\threedDII$ I also evaluated the path integral in spherical coordinates.
A detailed investigation of the kernel and wave-functions depend on the
particular choice of the boundary conditions (on $\threedDI$) and the
parameters $a,b$ (on $\threedDII$).
In order to achieve the results we had to incorporate a curvature term in the
definition of the quantum theory. Otherwise, a solution would not have been
possible. The particular form of the additional term was determined by the
method of (space-) time transformation. 
Of course, the solution of the Schr\"odinger equation by separation of
variables is also only possible with this additional curvature term.
This additional term can be cast in to
the form $(\hbar^2/2m)\times(R/8)$, where $R$ is the scalar curvature.
In view of the fact that our world has three spatial dimensions and
any theory of gravity requires spaces with (constant or non-constant)
curvature, we find the important feature that such models require curvature
terms in the corresponding Lagrangian in order to set up a proper and
solvable quantum theory. 

Therefore we found on $\threedDI$ seven separating coordinate systems, and on
$\threedDII$ the eleven coordinate systems of three-dimensional Euclidean
space. On  $\threedDI$ the $(u,v,w)$-system is singled out because the
cylindrical systems are contained as sub-systems, and the parabolic and 
paraboloidal systems cannot be evaluated further. The propagator on
$\threedDII$ can be evaluated also in other coordinate systems, in the 
remaining three cylindrical systems (as subsystems), and in parabolic
coordinates.   
 
I have not discussed the other two three-dimensional extensiuns of the Darboux
spaces as defined in \cite{KalninsKMWinter}. Their corresponding
generalization is more complicated and, furthermore, other coordinate systems
which separate the Schr\"odinger equations come into play, which appear on the
complex sphere and complex Euclidean space. This issue will be addresses in a
future publication. 

Having studied the free motion on these three-dimensional spaces, the 
next step is to search and investigate superintegrable potentials 
\cite{GROPOa,GROPOe,WSUF}. In particular, in three dimensions there is
a great variety of such potentials. In total, there are five maximally
superintegrable potentials \cite{GROPOa}, the first four of them also are
superintegrable on $\threedDII$, including the singular harmonic oscillator,
the Holt potential and the Coulomb potential. 
Studies along such lines are straightforward, many of the results from two
dimensions can be also used in the corresponding three-dimensional cases; this
will be investigated in a future publication.

\subsection*{\bf Acknowledgments}
The author is grateful to E. Kalnins for fruitful and pleasant
discussions on super-integrability and separating coordinate systems,
in particular for pointing out to me that the additional curvature term can be
cast into the form $\propto R/8$.

I would like to thank the organizers, in particular G.Pogosyan,  of the Second
International Workshop on Superintegrable Systems in Classical and Quantum
Mechanics for the warm hospitality during my stay at JINR, Dubna, Russia.

This work was supported by the Heisenberg-Landau program.


\newpage\noindent
\renewcommand{\baselinestretch}{0.90}


\end{document}